\documentstyle[bezier,12pt]{article}
\newcommand{\rf}[1]{(\ref{#1})}
\newcommand{\bea}{\begin{eqnarray}}
\newcommand{\eea}{\end{eqnarray}}
\newcommand{\g}{\gamma}
\renewcommand{\l}{\lambda}
\renewcommand{\b}{\beta}
\renewcommand{\a}{\alpha}

\newcommand{\m}{\mu}

\newcommand{\ep}{\varepsilon}
\newcommand{\om}{\omega}

\newcommand{\vph}{\varphi}
\newcommand{\oh}{\frac{1}{2}}
\newcommand{\oq}{\frac{1}{4}}

\newcommand{\dg}{\dagger}

\newcommand{\tr}{{\rm Tr}\;}

\newcommand{\vds}{v_{ds}}

\newcommand{\tH}{\tilde{H}}
\newcommand{\vt}{\tilde{v}}

\def\void{}
\def\labelmark{}

\newenvironment{formula}[1]{\def\labelname{#1}
\ifx\void\labelname\def\junk{\begin{displaymath}}
\else\def\junk{\begin{equation}\label{\labelname}}\fi\junk}%
{\ifx\void\labelname\def\junk{\end{displaymath}}
\else\def\junk{\end{equation}}\fi\junk\labelmark\def\labelname{}}

\newenvironment{formulas}[1]{\def\labelname{#1}
\ifx\void\labelname\def\junk{\begin{displaymath}\begin{array}{lll}}
\else\def\junk{\begin{equation}\label{\labelname}\left.
\begin{array}{lll}}\fi\junk}%
{\ifx\void\labelname\def\junk{\end{array}\end{displaymath}}
\else\def\junk{\end{array}\right.\end{equation}}
\fi\junk\labelmark\def\labelname{}\def\junk{}
}

\newcommand{\beq}{\begin{formula}}
\newcommand{\eeq}{\end{formula}}
\newcommand{\beqv}{\begin{formula}{}}

\begin{document}
\topmargin 0pt
\oddsidemargin 5mm
\headheight 0pt
\headsep 0pt
\topskip 9mm

\addtolength{\baselineskip}{0.20\baselineskip}
\hfill    NBI-HE-92-28

\hfill May 1992
\begin{center}

\vspace{36pt}
{\large \bf

Non-perturbative
2d quantum gravity and hamiltonians unbounded from below}

\vspace{36pt}

{\sl J. Ambj\o rn and C.F. Kristjansen}

\vspace{12pt}

 The Niels Bohr Institute\\
Blegdamsvej 17, DK-2100 Copenhagen \O , Denmark\\

\end{center}

\vspace{24pt}

\vfill

\begin{center}
{\bf Abstract}
\end{center}

\vspace{12pt}

\noindent
We show how the stochastic stabilization provides both the weak coupling
genus expansion and a strong coupling expansion of 2d quantum gravity.
The double scaling limit is described by a hamiltonian which is unbounded
from below, but which has a discrete spectrum.

\vspace{24pt}

\vfill

\newpage

\section{Introduction}

One of the problems of euclidean quantum gravity is that the action is not
bounded from below. One might think that this is not a serious problem in two
dimensions where the Einstein-Hilbert action is topological:
\beq{*Euler}
\int d^2 \xi \sqrt{g} R =2\pi \chi
\eeq
However, in case one attempts a summation over all genera, as {\it is} needed
in string theory and as {\it might}  be needed in quantum gravity, it
creates ambiguity. This ambiguity is highlighted by analyzing 2d gravity
in the matrix model approach. It is conveniently  formulated
as follows: In the regularized path integral the class of manifolds
to be included  can be chosen to be the  piecewise linear  orientable
manifolds, which one can get by gluing together  equilateral triangles. The
curvature assignment can then be done according to Regge calculus which
means that the curvature is assigned to the vertices and is proportional
to the number of triangles sharing a given vertex.
The gluing  of $K$ labeled oriented triangles can be done
in a systematic way by associating with
each oriented link a hermitian $N\times N$ matrix $\phi_{\a\b}$ (see fig.1)
\begin{center}
\begin{picture}(110,70)(1,50)

\put (1,50){$\a$}
\put (50,116){$\b$}
\put (95,50){$\g$}

\put (8,80){$\phi_{\a\b}$}
\put (75,80){$\phi_{\b\g}$}
\put (50,40){$\phi_{\g\a}$}

\put (10,50){\circle*{4}}
\put (10,50){\vector(2,3){20}}
\put (10,50){\line(2,3){40}}
%\put (10,50){\line(3,-1){60}}
%\put (10,50){\line(2,1){40}}
%\put (50,70){\line(2,-1){40}}
%\put (50,70){\circle*{4}}
%\put (50,110){\line(0,-1){40}}
\put (50,110){\circle*{4}}
\put (50,110){\vector(2,-3){20}}
\put (50,110){\line(2,-3){40}}
%\multiput(10,50)(8,0){10}{\line(1,0){4}}
\put (90,50){\vector(-1,0){40}}
\put (90,50){\line(-1,0){80}}
%\put (50,50){\circle*{4}}
\put (90,50){\circle*{4}}
\end{picture}

\vspace{8pt}

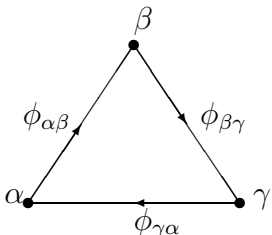
Figure 1: Labels of a triangle.

\end{center}
choosing an appropriate gaussian measure
\beq{*meas}
d\m_N(\phi) = c_{N} \exp(-\frac{1}{2} \sum_{\a,\b} |\phi_{\a\b}|^2)\;
\prod_{\a \leq \b} d\, {\rm  Re}\, \phi_{\a\b} \prod_{\a <\b} d \,{\rm Im}\,
\phi_{\a\b},
\eeq
and  taking the action to be proportional to the product of matrices
around the oriented triangle:
\beq{*act}
A(\phi) = \phi_{\a\b}\phi_{\b\g}\phi_{\g\a} ={\rm Tr}\,\phi^3.
\eeq
 In this way all closed triangulated surfaces of the kind mentioned
will be generated by the following field theoretical model
\beq{*tt1}
Z(h,N)=\int e^{h A(\phi)/\sqrt{N}} d\m_{N}(\phi).
\eeq
In fact, by expanding $e^{hA(\phi)}$, the Wick contractions of the term
$(h/\sqrt{N})^K A(\phi)^K/K !$
will generate all oriented 2D simplicial manifolds consisting of
$K$ triangles. Further the $1/N$-expansion organizes
the  surfaces according to powers of $N$.
For a surface with  Euler number  $\chi$
the total power of $N$ is $\chi$.

If we write the continuum euclidean action as
\beq{*cont}
S(\l,G)= \l \int d^2\xi \sqrt{g}-\frac{1}{2\pi G} \int d^2\xi \sqrt{g}R
\eeq
a  comparison with \rf{*Euler}
leads to the identification of $e^{\chi/G}$ and $N^\chi$:
\beq{*Gchi}
\frac{1}{G} = \log N
\eeq

Of course the field theory \rf{*tt1} is not well defined since the
action is not bounded from below. It should be emphasized that this
unboundedness has nothing to do with the cubic action. Had we decided
to glue together regular squares the action ${\rm Tr}\, \phi^3$ in \rf{*act}
would have been changed to a quartic action ${\rm Tr}\,  \phi^4$, but it
would appear with the wrong sign, since each surface should be assigned
a positive weight. However, each term
in the $1/N$-expansion is perfectly well defined and has a finite
radius of convergence.
The difference between the restricted summation provided by each term
in the $1/N$-expansion and the unrestricted summation over all surfaces
is the following: For a fixed topology, i.e. fixed $\chi$, the number of
surfaces which can be constructed by gluing together $K$ equilateral
triangles grows exponentially, while the total number of such surfaces grows
factorially, i.e. much faster:
\begin{formulas}{*entropy}
{\cal N}_\chi (K) &\sim & K^{\g_\chi -3} \; e^{\m_c K} \\
{\cal N}_{tot} (K) & \geq & {\rm const.}\;K !
\end{formulas}
The exponential bound in \rf{*entropy} leads to a critical point
which is the same for all $\chi$ and which allows us to take a
scaling limit for $\mu \to \m_c$ for each individual $\chi$.
 One gets
\beq{*88}
Z_\chi (\mu,G)= e^{\chi/G}\; \sum_K {\cal N}_\chi (K) e^{-\m K}
\sim \frac{c_\chi\;e^{\chi/G}}{(\m-\m_c)^{\g_\chi-2}}
\eeq
In \rf{*88} $\m$ is related to $h$ in \rf{*tt1} by
\beq{*hmu}
h=e^{-\m}
\eeq

The so-called double-scaling limit is an attempt
to go beyond this expansion,
and can be viewed as a formal renormalization of the gravitational
constant $G$ in front of the Einstein-Hilbert action.
The key observation is that
\beq{*key}
\g_\chi = -\frac{5}{4} \chi+2
\eeq
which allows us to write
\beq{*Ztotal}
Z (\m,G) \sim  \sum_\chi c_\chi \; e^{\chi/G_R}
\eeq
where the renormalized gravitational constant is defined as
\beq{*Grenorm}
\frac{1}{G_R} =\frac{1}{G}+\frac{5}{4} \log (\m-\m_c)
\eeq
The remarkable fact that the partition function in the scaling
limit is only a function of the renormalized gravitational
coupling constant was first observed in \cite{david,dk} in the
context of Liouville theory, and in the context of the matrix model
in \cite{ds,gm,bk}.

The factorial growth \rf{*entropy} of the total number of
piecewise linear surfaces is reflected in a factorial growth of
the coefficients $c_\chi$ in \rf{*Ztotal}. The series is only
an asymptotic series and not even Borel summable since all the
coefficients are positive. This is consistent with the
unboundedness of the action \rf{*act}. The origin of the unbounded
action \rf{*act} is therefore different from the usual problem caused
by the conformal mode in euclidean gravity, but it is a problem
which will be present even more severely in higher dimensional
gravity.

Two possible cures of  the unboundedness of the action in euclidian quantum
gravity were suggested long time ago. One method is based on stochastic
quantization and uses the bounded Fokker-Planck potential \cite{jeff}.
The other method involves a rotation of the integration contour in the
functional integral into the complex plane such that the functional
integral converges \cite{hawking}. They were both intended to be used
to cure the problem of the conformal mode, but have also been applied
to regularize the summation over topology in 2d gravity
\cite{mp,ajv,km,spanish,russian,david1}. The outcome is that the
two methods do not agree\cite{aj}. In fact the contour rotation leads
to a complex partition function and complex correlators, a signal that
either the method is no good or the theory itself is incomplete.
The stochastic method has no obvious flaws and in the following
will will analyze this method in detail\footnote{Here it should
also be mentioned that yet another suggestion of
a non-perturbative regularization can be found in the papers by Morris et al.
\cite{morris}. This regularization has no obvious flaws either, but
does unfortunately not agree with the stochastic method\cite{ajm}.
At the moment we do not have any convincing arguments in favor of one of the
the two methods.}.

\section{Stochastic regularization}

The stochastic quantization scheme for a $d$-dimensional euclidean field
theory with an action $S[\phi]$
shows that the vacuum expectation value of any operator $Q$ can be
interpreted as  the ground state expectation value in a
$(d+1)$-dimensional quantum theory:
\begin{formula}{*stoch}
<Q>= \frac{1}{Z} \int d \phi \; e^{-S[\phi]/\hbar} Q[\phi]=
\int d\phi \; \Psi_0^2 [\phi]\; Q[\phi].
\end{formula}
Here
\beq{*psi}
\Psi_0 [\phi] = \frac{e^{-S[\phi]/2\hbar}}{\sqrt{Z}}
\eeq
is the ground state of the theory, determined by the Fokker-Planck
hamiltonian
\beq{*FPH}
H_{FP} = \int d^d x \left[ -\frac{\delta^2}{\delta \phi^2}
+\frac{1}{4\hbar^2} \left(\frac{\delta S}{\delta \phi}\right)^2-
\frac{1}{2\hbar}\frac{\delta^2 S}{\delta\phi^2 } \right]
\eeq
This hamiltonian is a positive semidefinite operator and can be written
as
\beq{*RR}
H_{FP} = \int d^dx \; R^\dg (x) R(x),~~~~~~
R=i\frac{\delta}{\delta \phi}+\frac{i}{2\hbar} \frac{\delta S}{\delta \phi}
\eeq
and it is readily verified that \rf{*psi} is an eigenvector of $H_{FP}$
corresponding to energy $E=0$.

For this to make sense it has been assumed that  the action $S[\phi]$
 produces a normalizable wavefunctional by \rf{*psi}.
This is clearly not the case if $S[\phi]$ is unbounded from below.
Formally the wavefunctional \rf{*psi} will still satisfy $H_{FP} \Psi_0=0$,
even for a bottomless action, but it does not qualify as a groundstate
as it is not normalizable. The true ground state will correspond to the
lowest eigenvalue of $H_{FP}$ where the eigenstate is normalizable. If
we denote these by $E_0^{(T)}$ and (again) by $\Psi_0$, we can write in the
case
of a bottomless action:
\beq{*PsiN}
\Psi_0[\phi] = \frac{e^{-S_{eff}[\phi]/2\hbar}}{\sqrt{Z_{eff}}},
{}~~~~S_{eff} \neq S,~~~~E_0^{(T)} > 0.
\eeq
This equation can serve as a definition of a new stabilized action.
The expectation values of observables are defined as in \rf{*stoch}:
\beq{*expv}
\langle Q \rangle = \frac{1}{Z_{eff}} \int d\phi \;Q[\phi]\;
e^{-S_{eff}[\phi]/\hbar} \; =\; \langle \Psi_0 | Q | \Psi_0 \rangle
\eeq
and it can be shown \cite{jeff} that they have the same classical
limit, the same perturbative expansion in coupling constants
and the same  $1/N$ expansion as in the ill-defined bottomless theory.
Especially the last point will be important for us.

Let us now apply the above  formalism to the matrix models describing
2d-gravity. The partition function of 2d-gravity is in the matrix
model approach  given by
\beq{*2dp}
Z= \int d\phi e^{-N \tr V(\phi/\sqrt{N})}
\eeq
where $\phi$ denotes a $N\times N$ hermitean
matrix and the potential $V(x)$ is
characterized by starting with a quadratic term and then containing
a number of wrong sign higher power
terms $-g_3x^3-g_4x^4 -\cdots$. As explained
above the cubic term corresponds to the gluing of triangles, the quartic term
to the gluing of squares etc. The critical behaviour is universal
as long as the coupling constants $g_3,g_4,...$ are positive and is
uniquely determined by the {\it quadratic} nature of the maximum of
$V(x)$ for $x >0$. We will not consider the more general
situation, where the coupling constants $g_3,g_4,...$ can have different
signs. In that case a fine-tuning of the $g_i$'s with different signs will
result in different critical behaviour whenever the first extremum of
$V(x)$ for $x>0$ will be of higher order than two. Such critical behaviour
corresponds to non-unitary matter coupled to 2d-gravity. The non-unitarity
is not  unexpected in view of the surface representation, since it corresponds
to gluing certain polygons to the surface with negative weight.
We will here restrict ourselves to the simplest potential, which also
has the nicest interpretation in terms of geometry: the construction
of triangulated surfaces:
\beqv
V(\tilde{x})= \oh \tilde{x}^2 -g_3 \tilde{x}^3
\eeq
It is convenient by a simple translation and rescaling to
use it in the form
\beq{*pot}
V(x)= g x- \frac{x^3}{3}
\eeq
Exploiting the assumed $U(N)$ invariance of the vacuum we
follow the classical treatment in \cite{bipz} and diagonalize the
hamiltonian \rf{*FPH} with respect to the eigenvalues $x_i$, $i=1,..,N$
of the rescaled matrix $\vph=\phi/\sqrt{N}$
\begin{eqnarray}
H_{FP}[\phi] &=& N \sum_{i=1}^{N} H_{fp}[x_i] \label{h1}\\
H_{fp}[x] &=& -\frac{1}{N^2} \frac{d^2}{dx^2}+V_{fp}(x) \label{h2}\\
V_{fp}(x)&=& \oq  (g-x^2)^2+x \label{h3}
\end{eqnarray}
For the simple potential \rf{*pot} we see that $H_{FP}[\phi]$ just
becomes the sum of $N$ non-interacting single particle hamiltonians.
The particles behave as fermions since the expectation value of
any $U(N)$ invariant observable , calculated according to \rf{*expv}, is
\beq{*fvev}
\langle Q[\phi] \rangle =
\int \prod_{i=1}^{N} dx_i \, \Delta^2 (\{ x_i \}) \, \Psi_0^2(\{ x_i \})
Q(\{ x_i \})
\eeq
where $\Delta(\{ x_i \})$ comes from the integration over the
angular part of the $\phi$ variable, and is given by the Vandermonde
determinant:
\beq{*van}
\Delta (\{ x_i \}) = \prod_{i < j} (x_i-x_j)
\eeq
In this way, as already observed in \cite{bipz}, the function
\beq{*fer}
\Phi(\{ x_i \}) \equiv \Delta (\{ x_i \}) \, \Psi_0(\{ x_i \})
\eeq
becomes the totally antisymmetric groundstate wave function of a fermionic
system of $N$ non-interacting particles and can therefore be written
as the Slater determinant of the $N$ lowest single particle  eigenfunctions
of $H_{fp}[x]$:
\begin{equation}
\Phi(\{ x_i \}) = \frac{1}{\sqrt{N !}} \left|
\begin{array}{ccc}
\chi_1(x_1) & \cdots & \chi_N(x_1) \\
\vdots &   ~         &  \vdots \\
\chi_1(x_N) & \cdots  & \chi_N(x_N)
\end{array}                           \right|
\label{*det}
\end{equation}
where
\beq{*eigenv}
H_{fp} \chi_j(x) =E_j \chi_j(x)
\eeq
and where the ground state energy of the total system
is
\beq{*E0}
E_0^{(T)}= N (E_1+ \cdots+ E_N )
\eeq

In the context of the $1/N$-expansion the total energy $E_0^{(T)}$ is
zero. As explained above the ground state energy of $H_{FP}$
is trivially zero when the original problem is well defined and
\rf{*psi} is normalizable\footnote{This is a general property
of semi-definite operators of the form \rf{*RR}, as is well known
from supersymmetric quantum mechanics. However, we feel that supersymmetry
in this context is somewhat of a red heering. First, the supersymmetry
alluded to in the context of matrix models has not yet been given a
useful physical interpretation, and next the phenomenon is more general.
As an example we can mention the case of ordinary 4d Yang-Mills theory
where the operator $R(x)$ in temporal gauge is $i\delta/\delta A(x)-iB(x)$,
and an exact solution to the functional Schr\"{o}dinger operator
corresponding to $E_0=0$ is $\exp (-N_{cs}[A])$, where $N_{cs}[A]$ denotes the
Chern-Simons number of the gauge configuration. However, this solution
is not normalizable, as is the case for the true groundstate which has
$E_0>0$.}. Even if the original matrix problem looks
unbounded (it involves $\phi^3$ terms, wrong sign $\phi^4$ terms etc)
it is actually well defined to any finite order in the $1/N$ expansion
as long as the coupling constants are less than the critical coupling
constants which determine the radius of convergence of the sum of
diagrams to leading order in the $1/N$ expansion. This is reflected
in the fact that the distribution of eigenvalues of the matrices
to the leading order in $N$ is restricted to a finite interval.
The wave function, as given by \rf{*psi}, is therefore normalizable
in this leading approximation, and in fact to all finite orders
in $1/N$.  Using the stochastic regularization one is able to
go beyond the $1/N$ expansion and in addition one is able to
extend the theory beyond the critical value of the coupling constant.
When the  coupling constant is beyond the radius of
convergence  the matrix model is not defined even in the planar
approximation. But again stochastic regularization will provide
us with a definition of the theory in this region of coupling
constant space.

\section{The large-N limit}

Since the system of $N$ non-interacting particles is fermionic and since
we have to perform the summation over the first $N$ eigenvalues, let us
denote the energy of the $N$'th level by $E_F$ (the Fermi energy).
{}From \rf{h2} we see that large $N$ corresponds to the semiclassical
limit $\hbar = 1/N \to 0$. We expect therefore  the WKB approximation
to be good for all but the lowest eigenvalues, unless special
circumstances occur\footnote{We shall discuss these special
regions of coupling constant space later.}. According to the Bohr-Sommerfeld
quantization rule the phase space of the classical theory is related
to the $n$'th energy level of $H_{fp}$ by:
\beq{*I1}
2\pi n \hbar = \int dp\,dx\; \theta[ E_n-(p^2 + V_{fp}(x))]
\eeq
where $\hbar = 1/N$. This gives the following relation:
\beq{*I2}
\frac{n}{N} = \frac{1}{\pi} \int_{x_l(E_n)}^{x_r(E_n)}
dx\; \sqrt{E_n -V_{fp} (x)}
\eeq
where $x_r(E_n)$ and $x_l(E_n)$ are the classical turning points for
a particle with energy $E_n$ moving in the potential $V_{fp} (x)$. The
Fermi energy is then determined by
\beq{*I3}
1 = \frac{1}{\pi} \int_{x_l(E_F)}^{x_r(E_F)}  dx \;
\sqrt{E_F - V_{fp}(x)}
\eeq
while the energy density is
\beq{*I4}
\rho (E) \equiv \frac{\partial n}{\partial E} =
\frac{N}{2 \pi}  \int_{x_l(E)}^{x_r(E)}  dx \;
\frac{1}{\sqrt{E-V_{fp}(x)}}.
\eeq

\vspace{12pt}

We can now simplify the expression \rf{*fvev} if we restrict ourselves
to observables of the form
\beq{*I5}
\frac{1}{N} {\rm Tr}\, f(\phi) = \frac{1}{N} \sum_{i=1}^{N} f(x_i).
\eeq
By expanding the Slater determinant and using the orthogonality of the
single particle wave functions $\chi_n$ we get
\beq{*I6}
\langle \frac{1}{N} {\rm Tr}\, f(\phi)\rangle = \int dx \; u(x) f(x)
\eeq
\beq{*I7}
u(x) \equiv \frac{1}{N} \sum_{i=1}^{N} \chi_i^2 (x)~;~~~~~
\int_{-\infty}^{\infty} dx \; u(x) =1.
\eeq

The density $u(x)$ can be calculated in the WKB approximation, since the
eigenfunctions $\chi_n(x)$ in this approximation in the classically allowed
region where the energy $E_n > V_{fp}(x)$ are given by
\beq{*I8}
\chi_n(x) \approx \frac{C(E_n)}{(E_n-V_{fp}(x))^{1/4}}
\cos \left(N \int^x dy \sqrt{ E_n-V_{fp} (y)} \right)
\eeq
The normalization is fixed by
\beq{*I8a}
\int_{-\infty}^{\infty} dx \; \chi^2 (x) =1
\eeq
where the integration is effectively cut off at the classical classical
turning points $x_l(E_n)$ and $x_r(E_n)$. Due to Riemann's lemma the
$\cos^2$ is replaced by $1/2$ in the large $N$ limit and
\beq{*I9}
\frac{1}{C^2(E_n)} = \oh \int_{x_l(E_n)}^{x_r(E_n)}  \;
\frac{dx}{\sqrt{ E_n-V_{fp}(x)}} = \frac{\pi}{N} \rho (E_n)
\eeq
We conclude that $u(x)$ in the large $N$ limit is
\bea
u(x) &\equiv & \frac{1}{N} \sum_{i=1}^{N} \chi_i^2(x) \nonumber \\
{}~ & = & \frac{1}{2\pi} \int_{V_{fp}(x)}^{E_F} dE
\frac{1}{\sqrt{ E-V_{fp}(x)}} = \frac{1}{\pi} \sqrt{E_F -V_{fp}(x)}
\label{*I10}
\eea

It is clear from \rf{*I6} that $u(x)$ should be given the interpretation
of density of eigenvalues in the original matrix model. Formulae
\rf{*I6}-\rf{*I10} give the corresponding derivation in the context of
stochastic regularization and we see that the finite range of
eigenvalues in the large $N$ limit is determined by the classical
turning points corresponding to the Fermi energy $E_F$.

Although we know from the general theorems \cite{jeff} that $u(x)$ has to
agree with the eigenvalue density in the original 0-dimensional matrix
model, it is interesting to see how this comes about\footnote{This argument
was first presented in \cite{spanish}.}. From the classical work of
Brezin {\it et al.} \cite{bipz} we know that the semiclassical
eigenvalue density of a matrix model described by a potential
$V(x)$ is given by
\beq{*I11}
u(x) = \frac{1}{\pi} \sqrt{ \frac{1}{k-1} V''(x)-\frac{1}{4} (V'(x))^2+
P_{k-3}(x)}
\eeq
if $V(x)$ is a polynomial of order $k$. In \rf{*I11} the polynomial
$P_{k-3}(x)$ of order $k-3$ is fixed by requiring the support of $u(x)$
to be connected, In the case where $k=3$ we get
\beqv
u(x)= \frac{1}{\pi} \sqrt{ P_0 - V_{fp} (x)}
\eeq
and the  constant $P_0$ is fixed by the normalization condition for $u(x)$.
A glance on \rf{*I3} allows us to identify $P_0$ with the Fermi energy in
the potential $V_{fp}(x)$.

\vspace{12pt}

\begin{center}
\setlength{\unitlength}{0.1cm}

\begin{picture}(150,80)(0,0)

\bezier{500}(5,75)(10,5)(25,5)
\bezier{500}(25,5)(35,5)(41,35)
\bezier{500}(41,35)(45,55)(50,55)
\bezier{500}(50,55)(53,55)(55,50)
\bezier{500}(55,50)(58,45)(60,45)
\bezier{500}(60,45)(65,45)(70,75)

\bezier{500}(90,75)(95,5)(110,5)
\bezier{500}(110,5)(120,5)(126,35)
\bezier{500}(126,35)(130,55)(137,57)
\bezier{500}(137,57)(145,59)(148,75)

\multiput(9,45)(4,0){13}{\line(1,0){2}}
%\put(5,45){\line(1,0){65}}
\multiput(93,57)(4,0){11}{\line(1,0){2}}
%\put(90,57){\line(1,0){60}}

\end{picture}

\end{center}

\vspace{8pt}

\noindent
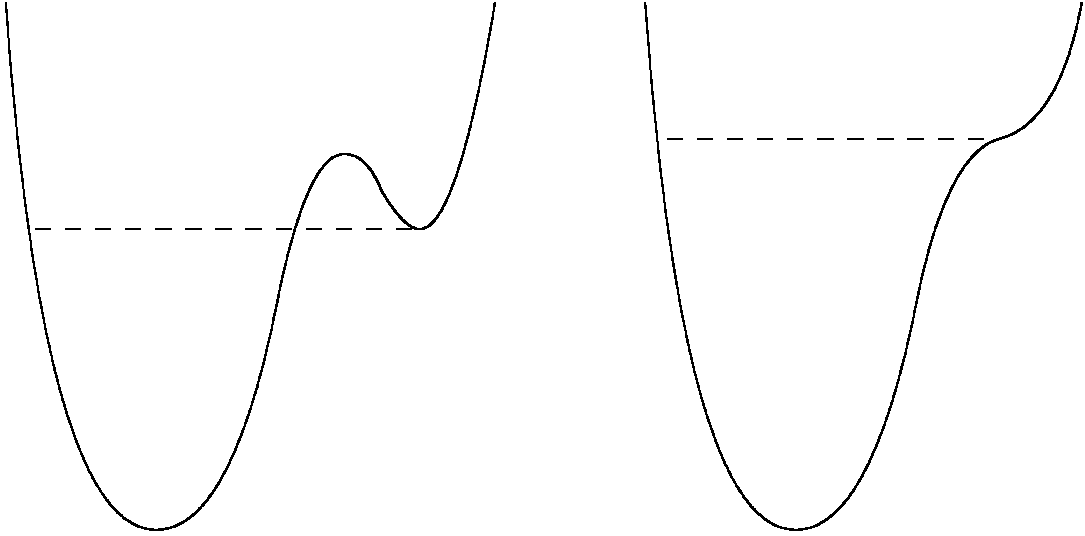
Fig.2 The figure to the left (fig.2a) has $g > g_c$. For the other figure
(fig.2b) $g < g_c$. The dashed lines represent the  Fermi energy in
the two cases.

\vspace{12pt}

If we return for a moment to the original matrix model given by the
potential $V(x)= gx-x^3/3$ the sum over planar diagrams which constitutes the
large $N$ limit has a finite radius of convergence as a power series
in the coupling constant $(1/g)^{3/4}$, as explained in the
introduction\footnote{ It is a power series in $(1/g)^{3/4}$
rather than $g$ due
to the rescaling and translation $ \tilde{x}^2- \tilde{g} \tilde{x}^3
\to gx-x^3/3$.}. This radius is determined from the eigenvalue distribution
$u(x)$. The power series is convergent as long as the function
$\sqrt{V_{fp}(z)-E_F}$, $z$ complex, has a single cut on the real axis
\cite{bipz}. This is only possible if the fourth order polynomial
$V_{fp}(x)$ has the form shown in fig. 2a and if $E_F$ coincides with
the second minimum of the potential, since in this case
\beq{*I12}
V_{fp}(z)-E_F = \frac{1}{4} (z-x_+)^2 (z-x_r)(z-x_l)
\eeq
where $x_+$ denotes the position of the local minimum at the far right,
while $x_l$ and $x_r$ are the classical turning points in $V_{fp}$ for
the energy $E_F$. {\it  It is a remarkable fact that $E_F$ coincides with
the local minimum at $x_+$} \cite{mp}. From fig.2a it is clear that there
will be a critical value $g_c$ below which \rf{*I12} cannot be realized
and where $x_+$ will in fact split in two complex conjugate zeroes. For
this value of $g_c$~ $x_+$ and $x_r$ merge to a single point $x_c$ where
\beq{*I13}
V_x'(x_c,g_c) =0,~~~~~~V_{xx}''(x_c,g_c)=0.
\eeq
We find from \rf{*I13}
\beq{*I14}
g_c = \frac{3}{2^{2/3}},~~~~~~x_c= \sqrt{ \frac{g_c}{3}}.
\eeq
The planar expansion will be convergent for $g>g_c$ and the eigenvalue
distribution for a given $g>g_c$ will be located between $x_l$ and $x_r$.
This is true to any finite order in the $1/N$ expansion. Note
however that while the original 0-dimensional matrix model made no sense
below $g_c$ the stochastically regularized model is perfectly well defined
even in this range of coupling constants in agreement with general
results \cite{jeff}.

\section{The double scaling limit}

Let us consider the situation where $g > g_c$ (fig. 2a) and show how the
double scaling limit as a function of only one parameter,
\beq{*I15}
\hbar^2= \frac{4 g_c^{5/2}}{N^2 (g-g_c)^{5/2}}
\eeq
appears in a trivial way as the WKB expansion of the one-particle
hamiltonian $H_{fp}$.

If we introduce a scaled variable
\beq{*I16}
z= \sqrt{\frac{g_c}{g-g_c}}\; \frac{x-x_c}{x_c}
\eeq
the Fokker-Planck hamiltonian \rf{h2} can be written as
\beq{*I17}
H_{fp} = V_{fp}(0;g-g_c)+ x_c^4 \frac{(g-g_c)^{3/2}}{g_c^{3/2}}\;
h_{fp}(z)
\eeq
\beq{*I18}
h_{fp}(z) = -\hbar^2 \frac{d^2}{dz^2}+ v_{fp}(z;\sqrt{g-g_c})
\eeq
\bea
v_{fp}(z;\sqrt{g-g_c})& = &
-3z+z^3 + \frac{\sqrt{g-g_c}}{\sqrt{g_c}}\,
\left[-\frac{3}{2} z^2 + \frac{1}{4} z^4 \right] \nonumber\\
{}~&=& v_{ds}(z) + o(\sqrt{g-g_c})  \label{*I19}
\eea
The constant $V_{fp}(0;g-g_c)$ is a second order polynomial in $g-g_c$.
The double scaling limit is defined as $N \to \infty$, $ g \to g_c$,
$\hbar$ fixed, and we see that the physics in this limit is determined
by the hamiltonian
\beq{*I20}
h_{ds} (z) = -\hbar^2 \frac{d^2}{dz^2}  + v_{ds} (z)
\eeq
To be precise we have dropped the trivial analytic behaviour in
$g-g_c$ present in \rf{*I17} in the term $V_{fp}(0;g-g_c)$. Likewise
one has to extract a power of $g-g_c$ in order to get a non-trivial
result in the limit $g \to g_c$, as is well known from the conventional
analysis of the 0-dimensional matrix models.

The hamiltonian \rf{*I20} admits a WKB expansion in $\hbar^2$.
The fact that 0-dimensional matrix
models allow for such an expansion was a non-trivial result in the original
formulation (as well as in the continuum approach using Liouville theory),
but by the method of stochastic regularization we get it almost for free,
as was also the case for the eigenvalue distribution and the critical
point in the large $N$ limit.

Since the expansion in the double scaling limit is identical to the
WKB expansion of \rf{*I20} it is worth while to formulate explicitly
the {\it exact} WKB equation corresponding to \rf{*I20}. The leading order
was already discussed above, but we can write in general:
\bea
\chi (z) &=& A(z) e^{\pm iS(z)/\hbar}     \label{*I21} \\
A(z)   & =& A^{(0)}(z) + \hbar^2 A^{(1)}(z) + \cdots  \label{*I22}  \\
S(z)  & = & S^{(0)} (z) + \hbar^2 S^{(1)} (z) + \cdots  \label{*I23}
\eea
and the Schr\"{o}dinger equation
\beq{*I24}
h_{ds}(z) \chi_e(z) = e \chi_e (z)
\eeq
is {\it equivalent} to  the exact WKB equations:
\beq{*I25}
S' = c/A^2
\eeq
\beq{*I26}
-\hbar^2 A'' A^3 + c^2 = (e-v_{ds}(z))A^4.
\eeq
The constant $c$ in \rf{*I25} is determined by the normalization
of $\chi_e$. If we introduce $R=A^2$ \rf{*I26} can be written as
\beq{*I27}
-\hbar^2 \left( R_{e}''(z) R_e (z) - \frac{1}{2} R_{e}'(z)^2 \right)+2c^2
= 2(e-v_{ds}(z))R^2_{e} (z)
\eeq
or differentiating in order to get rid of c, if wanted:
\beq{*I28}
-\frac{\hbar^2}{2} R_{e}'''(z) + v_{ds}'(z)R_e (z) =
2(e-v_{ds} (z))R_{e}'(z).
\eeq
This is the so-called non-perturbative equation first derived in \cite{km}
by means of the Dikii-Gelfand equation. We see that it is nothing but the
WKB equation for $A_e^2(z)$ and in our opinion it seems to have no
advantage compared to the original Schr\"{o}dinger equation   \rf{*I24}.

\vspace{12pt}

At first sight it appears as if we have gained nothing if we take the
double scaling limit as in \rf{*I20}. The problem
of dealing with the original unbounded action has been replaced by the
problem of how to deal with the hamiltonian \rf{*I20}, which is unbounded
from below. However, hamiltonians unbounded from below are perfectly
respectable objects from a mathematical point of view, as we will
review in the next section. They have not played a significant role
in quantum mechanics, due to the lack of situations where they
appear in a natural way. It is somewhat paradoxical that one
has to go all the way to quantum gravity to find such a situation.

\section{Hamiltonians unbounded from below}

Consider a hamiltonian like \rf{*I20}, where the potential is unbounded
from below. To be more precise we will write
\beq{*h}
h= -\frac{d^2}{dy^2} +v(y)
\eeq
where we assume that  $v(y) \to -\infty$ for $y \to -\infty$ so fast that
\beq{*infty}
\int_{-\infty}^0 \frac{dy}{\sqrt{|v(y)|}} < \infty
\eeq
This condition means that a classical particle, once it is not trapped
in local energy minima of the potential $v (y)$, will move to
$y=-\infty$ {\it in a finite time} since a classical particle with
energy $e$ and hamiltonian $p^2+v(y)$will move from $y_t$ to $-\infty$ in time
\beq{*T}
T= \int_{-\infty}^{y_t} \frac{dy}{\sqrt{e-v(y)}}
\eeq
provided $e-v(y) >0$ for all $y \leq y_t$. The dynamical problem is not
{\it classical complete}, in the sense that we need to specify some
boundary conditions at $y=-\infty$ if we want to be able to address
dynamical questions ranging over all times. From this point of view the
mathematical aspects of the situation are not much different from the
considerations of dynamics in a finite box. The situation transfers to
the quantum mechanical case as well. Had \rf{*h} been defined in a finite
box we would have been forced to impose a boundary condition for each wall in
the box in order to get a self-adjoint operator. For each wall there
would be a one-parameter family of self-adjoint extensions of the symmetric
operator defined by \rf{*h} and acting on functions with support which does
not include the coordinate of the wall itself.  The spectral theory of
such operators is well known. In the case of two walls and no singularities
of $v(y)$ the spectrum is purely discrete and extends to infinity for
any of the self-adjoint extensions. What is less known to physicists is that
this result extends to potentials unbounded from below and satisfying
\rf{*infty}.  In fact the usual situation of walls with imposed boundary
conditions can be treated as a special case of the potentials unbounded
from below, and although this might seem
somewhat perverse from the point of view of physics, it is not unnatural from
a mathematical point of view, where both situations can be classified as
Weil's circle-limit case of the Sturm-Liouville theory of second order
differential operators (see for instance \cite{germans} for a recent
discussion).  The important conclusion is that the spectra
of the self-adjoint extensions of \rf{*h} are purely discrete and extend
from $-\infty$ to $+\infty$, provided \rf{*infty} is satisfied and the
potential behaves in a similar way for $y \to +\infty$ or $v(y) \to
\infty$ for $y \to \infty$.

It is not the purpose here to describe in any detail the Sturm-Liouville
theory for potentials unbounded from below, but the important points
are easily explained. Suppose we want to solve the differential
equation
\beq{*1}
h \psi = e \psi
\eeq
where $h$ is given by \rf{*h} and \rf{*infty}. Far to the left the
potential goes to $-\infty$ and the WKB-approximation becomes an
excellent approximation. In this region we can write:
\beq{*2}
\psi(y;e,\a) \sim \frac{1}{(e-v(y))^{1/4}} \cos \left\{ \left(
\int^{y_t(e)}_{y}
\sqrt{e-v(x)}\;dx + \a \right) \right\}
\eeq
where $\a$ is an arbitrary angle and $y_t(e)$ denotes the classical
turning point for a particle coming from the left
(we assume that $e$ is so small
that this point exists).
We see that {\it any} solution
to \rf{*1}  is square-integrable at $-\infty$ thanks to the condition
\rf{*infty}. However,  this class of functions is too large
to constitute the domain of a self-adjoint version of $h$ since a
partial integration leads to a term:
\beq{*3}
 \psi_1^* (y;e_1,\a_1) \frac{d \psi_2(y;e_2,\a_2)}{dy} -
\frac{ d\psi^*_1 (y;e_1,\a_1)}{dy} \psi_2(y;e_2,\a_2)
\eeq
which does not go to zero  for $y \to -\infty$.
  Special choices of $\a$ ensure the convergence, namely
\beq{*alpha}
\a (e) = -\int_{-\infty}^{y_t(e)}\, dy \;
\left[ \sqrt{e-v(y)}-\sqrt{e_0-v(y)} \right]
+\int^{y_t(e_0)}_{y_t(e)}\sqrt{e_0-v(y)}
\eeq
where $e_0 \geq e$ (and where we again assume that $e_0$ is chosen such that
$y_t(e_0)$ exists). We see that the integrals in
\rf{*alpha}  give a well defined way to
write the formal difference in WKB
phases corresponding to energies $e$ and $e_0$:
\beq{*alpha1}
\a (e) = \int_{-\infty}^{y_t(e_0)}\, dy \;\sqrt{e_0-v(y)}
           -\int_{-\infty}^{y_t(e)}\, dy \;\sqrt{e-v(y)}
\eeq
With this choice  all functions in \rf{*2} have the asymptotic behaviour
\beq{*4}
\psi_{e_0} (y) \sim   \frac{1}{(-v(y))^{1/4}} \cos \left\{
 \left( \int^{y_t(e_0)}_{y} \sqrt{e_0-v(x)}\;dx +o((e_0-e)y/\sqrt{-v(y)})
 \right) \right\}
\eeq
and for  fixed $e_0$ \rf{*3} goes to zero for $y \to -\infty$.
The different self-adjoint extensions will be characterized by different
choices of the parameter $e_0$. Due to the asymptotic behaviour
\rf{*4} of the wave functions it is possible to characterize the
unbounded hamiltonian as a sequence of ordinary bounded hamiltonians
\cite{fahri}. For $-y$ sufficiently large all wave functions
$\psi(y;e,e_0)$ will vanish
at points $y_n$ given by:
\beq{*4a}
\int_{y_n}^{y_t(e_0)} dy\; \sqrt{e_0-v(y)} = \pi (n+1/2) +
o\left( \frac{(e_0-e)y_n}{\sqrt{-v(y_n)}}\right)
\eeq
Consequently one will get the same result if one replaces the
unbounded potential by a potential $v_n (y)$ cut off by an infinitely high
wall at $y_n$, provided $|e|$ is not too large. In this way the
self-adjoint hamiltonian, unbounded from below and
characterized by the parameter $e_0$, is
reached by a sequence of ordinary  hamiltonians, bounded from below,
and corresponding to the potentials $v_n(y)$.

This discussion, based on the WKB approximation, can be made
mathematical rigorous, but it is worth emphasizing that it can
be formulated independently of the WKB expansion.
Let $e$ be given and let $\psi_1(y;e)$ and $\psi_2(y;e)$ be
two independent solutions to \rf{*1} with a wronskian
(which of course is independent of $y$)
\beq{*5}
w = \psi_2'(y) \psi_1(y)-\psi_2(y)\psi_1'(y).
\eeq
If we define a class of functions by
\bea
f(y) &=& \psi_1 (y;e)\left\{c_1 + \frac{1}{w}
\int_{-\infty}^{y} dx \;\psi_2(x;e) g(x) \right\}+\nonumber\\
& &\psi_2(y;e)\left\{c_2 - \frac{1}{w}
\int_{-\infty}^{y} dx\; \psi_1(x;e) g(x)\right\}  \label{*6}
\eea
 where $g$ is any square-integrable function, it is readily seen that
\beq{*7}
hf =ef+ g~~~~{\rm and}~~~~f(y) \to c_1 \psi_1(y;e)+c_2\psi_2(y;e)~~~
{\rm for}~~~y \to -\infty.
\eeq
Since $f$ is square-integrable, $h$ maps $f$ into a square-integrable
function, and
using the asymptotic behaviour \rf{*7} in \rf{*3} it is seen that
the class of functions defined by \rf{*6} constitutes the domain
of a self-adjoint version of $h$ provided the socalled (complex) limit numbers
$c_1$ and $c_2$ satisfy
\beq{*8}
c_1 \cos \a_0 +  c_2 \sin \a_0 =0
\eeq
for some real $\a_0$. In this way we recover the WKB results above.

If we assume that the potential also satisfies a condition like
\rf{*infty} for $y \to \infty$ we can repeat the discussion above
and get a relation like \rf{*8}, just with other constants $c'_1,c'_2$ and
$\a'_0$.  In order to solve the eigenvalue equation we have to choose
$g$ equal to zero in \rf{*6} and we get  a matching condition at
(say) $y=0$ for $f$ and $f'$.
For fixed $\a_0$ and $\a'_0$ the matching can only be satisfied for
certain discrete values of $e$, leading to a quantization of the
energy eigenvalues in much the same way as for a particle in a box
of finite width.
If the potential instead goes to $+\infty$ for $y \to \infty$ we also get
a purely discrete eigenvalue spectrum. The solutions to
\rf{*1} will in this case  consist of exponentially  growing and
exponentially decaying parts for $y \to \infty$. The linear combination of
$\psi_1(y;e)$ and $\psi_2(y;e)$ dictated by \rf{*8} will
in general contain an exponentially growing part. Only for special discrete
values of $e$ will the wave function be exponentially decaying and
therefore square-integrable for $y \to \infty$. This last situation will
be the one of interest to us.

\section{Application to $v_{ds}(z)= -3z+z^3$ }

Let us apply the formalism of unbounded hamiltonians to the
hamiltonian \rf{*I20} and compare the results with similar results
obtained by using the full Fokker-Planck potential.
We first discuss the quantization of energies.
The potential has a  local maximum at $z=-1$ and a local minimum
at $z=+1$ with a value $v_{ds}(z=+1) = -2$. The same value of
the potential is obtained if $z=-2$, which therefore is the classical
turning point when the energy coincides with the local minimum,
as is the case for the  Fermi energy in the WKB limit for the full
Fokker-Planck potential $v_{fp}(z)$ given in \rf{*I19}. Let us
denote the Fermi energy of the full Fokker-Planck hamiltonian
\rf{*I18} by $e_f$. It is trivially related to the $E_F$
defined previously by
\beq{*J1}
E_F = V_{fp}(0;g-g_c) + x_c^4\,\frac{(g-g_c)^{3/2}}{g_c^{3/2}}\, e_f.
\eeq

As mentioned above the spectrum for
$h_{ds}(z)$ is purely discrete. If $\hbar$ is small and $e \leq e_f$
we know that
the solutions to the Schr\"{o}dinger equation are  well approximated by
the WKB solutions  \rf{*2} to the left of the classical turning point
$z_t(e)$.
To the right of the turning point it is given by similar expressions
which fall off or increase exponentially. The WKB matching condition
to the exponentially decreasing solution is
\beq{*J2}
\psi(z;e) \sim \frac{1}{(e-\vds(z))^{1/4}} \cos
\left\{  \frac{1}{\hbar} \int^{z_t(e)}_{z}
\sqrt{e-\vds(z')}\;dz' -\pi/4 \right\}
\eeq
and the quantization of energies in the WKB expansion
comes about by comparing with the phase requirement \rf{*alpha}
which was needed in order that we had a self-adjoint hamiltonian.

If we let $e'_0$, the parameter characterizing the self-adjoint
extension in the WKB language (not an
eigenvalue), be of order one, we get for the $n'$th eigenvalue below $e'_0$,
$e_n$:
\beq{*J3}
-\pi \hbar(n+3/4) = \int^{z_t(e_n)}_{-\infty}  dz
\; \left( \sqrt{e_n - \vds (z)}-\sqrt{e'_0-\vds (z)}\right)
-\int_{z_t(e_n)}^{z_t(e'_0)} dz \sqrt{e'_0-\vds (z)}.
\eeq
We can solve this equation in two regions
(assuming as usual that $\hbar$ is small).
Assume first that $e'_0-e_n$ is
of the order of $\hbar$, and denote by $e_0$ the
first eigenvalue below $e'_0$. We get:
\beq{*J3a}
e_n = e_0- 2\hbar \om n +o\left( (\hbar n )^{3/2}\right)
\eeq
where
\beq{*J3b}
\om = \pi/ \int_{-\infty}^{z_t (e_0)} dz \frac{1}{\sqrt{e_0-\vds(z)}}
\eeq
has the interpretation as the cyclic frequency in the
classical motion out to infinity in a finite time (see \rf{*T}).
The only arbitrariness present at the energy levels close to
the Fermi energy is therefore an all-over displacement of
the energy levels. If we choose $e_0'$ such that $e_0 =e_f$,
the Fermi energy, we get
\beq{*omegaf}
\om \equiv \om_f = \sqrt{3},
\eeq
which is in agreement with the similar calculation
done in the full Fokker-Planck potential:
\beq{*J3c}
e_n = e_f - 2\hbar \om_f n +o\left( (\hbar n)^{3/2}\right)
+ o\left((g-g_c)^{1/4}\right).
\eeq
The main source to the difference between \rf{*J3a} and \rf{*J3c}
is the difference in escape time to infinity  in the potential
$\vds(z)$ compared to the
time it takes to reach the left turning point in the Fokker-Planck
potential $v_{fp}(z)$. This difference is of the order $(g-g_c)^{1/4}$
and vanishes in the double scaling limit.

The above results  are even more manifest if we calculate the energy
density for small $\hbar$ where we
expect \rf{*J3} to be a good approximation and where the eigenvalues
are dense
\beq{*J4}
\frac{d n_{ds}}{d |e|} = \frac{1}{2 \pi \hbar} \int^{z_t(e)}_{-\infty}
\; \frac{dz}{\sqrt{e-\vds(z)}}.
\eeq
We see that all reference to  $e'_0$ drops out.
Let us compare this result with the corresponding result for the
full potential $v_{fp}(y)$. From \rf{*I4} we get by simple rescaling:
\beq{*J5}
\frac{d n_{fp}}{d |e|} = \frac{1}{2 \pi \hbar} \int^{z_r(e)}_{z_l(e)}
\; \frac{dz}{\sqrt{e-v_{fp}(z)}}.
\eeq
where $z_r(e)$ and $z_l(e)$ denote the right and left turning points.
We have $z_r(e) = z_t(e)+ o( \sqrt{g-g_c})$
and one can check that
\beq{*J5a}
\hbar \frac{d n_{fp}}{d e}=\hbar \frac{d n_{ds}}{d e} +o((g-g_c)^{1/4})
\eeq
provided $e$ is not too close to the bottom of $v_{fp} (z)$, i.e. provided
$e >> -(g-g_c)^{-3/2}$. Again we see that the difference between
the eigenvalue densities of $h_{ds}$ and $h_{fp}$ vanishes in the
double scaling limit as $(g-g_c)^{1/4}$, and that this is valid
for a much larger energy range than indicated by \rf{*J3a} and \rf{*J3c}.

The other region where we can solve  \rf{*J3} is for $-e_n >>
-e_f \approx 2$. In this region we get
\beq{*J5b}
e_n = -\left( \frac{\hbar n}{c} \right)^{6/5}\;
\left(1+ o( (n\hbar)^{-4/5})\right)
\eeq
where the constant $c$ is
\beq{*J5c}
c = \int^{-1}_{-\infty} \left( \sqrt{-y^3}- \sqrt{-1-y^3}\right)
+ \int^0_{-1} \sqrt{-y^3}
\eeq
We can write \rf{*J5b} as
\beq{*J5d}
e_n \approx -\frac{1}{c^{6/5}}\; \left( \frac{n}{N} \right)^{6/5} \;
\frac{1}{(g-g_c)^{3/2}}
\eeq
and even in the case of the the Fokker-Planck potential
will this expression for the energy  be valid
all the way down to the $N$'th level below the Fermi energy.
However, the sum of these large negative eigenvalues does
not contribute to the non-trivial scaling since we have, using \rf{*J5d}
\beq{*J5e}
\Delta E \sim N(g-g_c)^{3/2} \sum^{N} e_n \sim N^2.
\eeq

\vspace{12pt}

\noindent
Although the energy eigenvalues of the self-adjoint extensions agree with the
eigenvalues of the full Fokker Planck potential in the double scaling
limit up to an all-over displacement we need reference to the
full Fokker Planck potential in two ways if we want to calculate
physical observables like $\langle \tr \phi \rangle$ or more generally
\rf{*I5}.
We need it to define the top level (i.e. the Fermi energy)
and the bottom level in the
summation \rf{*I7}, but we also need it to define
a cut-off for large negative $z$ when we calculate expectation values
of operators like $\langle \tr \phi \rangle$. The reason is that although
the density $u(z)$ defined from \rf{*I7} is integrable out to $z=-\infty$
$u_{fp}(z)$ and $u_{ds}(z)$ will differ when we go beyond the
left turning point of for the full Fokker Planck potential. Beyond the left
turning point,
\beq{*tur}
z_l(e) \approx  -\frac{4 \sqrt{g_c}}{\sqrt{g-g_c}}
\eeq
$u_{fp}(z)$ falls of exponentially. It even vanish beyond $z_l(e)$ in the
large $N$ limit. This is not the case for $u_{ds}(z)$ and integrals
like $\int_{\Lambda} z^n u_{ds}(z)$ diverge for $\Lambda \to -\infty$.
If we introduce a cut-off
\beq{*cut}
\Lambda \sim -\frac{1}{\sqrt{g-g_c}}
\eeq
we get however the correct scaling behaviour of our observables. The only
features we need to borrow from the full Fokker Planck potential in order
to get the correct scaling behaviour of observables in the double
scaling limit is thus a Fermi energy of order one below
the local maximum of $v_{ds}(z)$ and a cut-off $\Lambda$ in the negative $z$
as given by \rf{*cut}. Note that such a cut-off by an infinite wall is by
no means  unnatural for the self-adjoint extensions since their
wave functions converge to the {\it same} oscillating function \rf{*4},
independent of the eigenvalues $e_n$, for $z \to -\infty$. We can now
put the wall at one of the zeros of the oscillating function \rf{*4}
and this will not influence the spectrum or eigenfunctions.
If we are only interested in the non-trivial scaling behaviour of
the physical observables which is associated with the asymptotic
WKB expansion in $\hbar^2$ all of the self-adoint extensions can be used.
If we however are interested in the full non-perturbative contributions
which survive in the double scaling limit, but which cannot be expressed
as powers of $\hbar^2$, the different self-adjoint extensions will differ,
and we have to use either the full Fokker Planck potential or the
unique self-adjoint extension which has an energy eigenvalue exactly
equal to the Fermi energy $e_f$ of the full Fokker Planck potential.
This will be clear in a moment when we calculate the lowest non-perturbative
correction in the limit of  small $\hbar$.

\vspace{12pt}

\noindent
Let us finally address the non-perturbative corrections which are not to be
be found as powers of $\hbar$ within a systematic WKB expansion.
When $g > g_c$ we have the situation in fig.2a. We have two wells.
A large one (unbounded from below in the double scaling limit) and
a smaller one above the Fermi energy to the right.
In the limit of small $\hbar$ the tunneling through the barrier will
be exponentially suppressed as $e^{-\Gamma/\hbar}$, where $\Gamma$ will be
calculated below,  and to a first
approximation we have independent states in the left well and the right
well. However, {\it the situation is remarkably similar to that of
a symmetric well, even if there at first sight is nothing symmetric
about the situation}. First we note that the level spacing is the
{\it same} on both sides of the barrier.  From \rf{*J3b} and \rf{*omegaf} we
get
that the level spacing in the left well is given by $2\hbar \om_f$
where $\om_f = \sqrt{3}$. The level spacing in the right well is
determined by expanding $h_{ds}(z)$ around the local minimum at $z=1$ and
we get with $\tilde{z}=z -1$:
\beq{*min}
h_{ds}(\tilde{z}) = -\hbar^2 \frac{d^2}{d \tilde{z}^2} +
3 \tilde{z}^2 + \tilde{z}^3
\eeq
and the harmonic frequency which determines the lowest lying levels in
the semiclassical approximation is just $\om = \sqrt{3}$.  Not only
is  the level spacing the same on both sides of the  barrier, but in addition
{\it the Fermi energy  in the well to the left coincides precisely
with the lowest energy in the well to the right}.

\vspace{12pt}

\begin{center}
\setlength{\unitlength}{0.1cm}

\begin{picture}(150,70)(0,20)

\bezier{500}(50,20)(60,70)(80,70)
\bezier{500}(80,70)(85,70)(90,60)
\bezier{500}(90,60)(95,50)(100,50)
\bezier{500}(100,50)(107.5,50)(115,90)

\multiput(62,55)(4,0){8}{\line(1,0){2}}
\multiput(106.5,55)(4,0){3}{\line(1,0){2}}
\multiput(58,45)(4,0){15}{\line(1,0){2}}
\multiput(54,35)(4,0){16}{\line(1,0){2}}
\put(20,55){\line(1,0){40}}
\put(94,55){\line(1,0){11}}

\put(20,45){\line(1,0){36}}
\put(20,35){\line(1,0){32}}

\end{picture}

\end{center}

\noindent
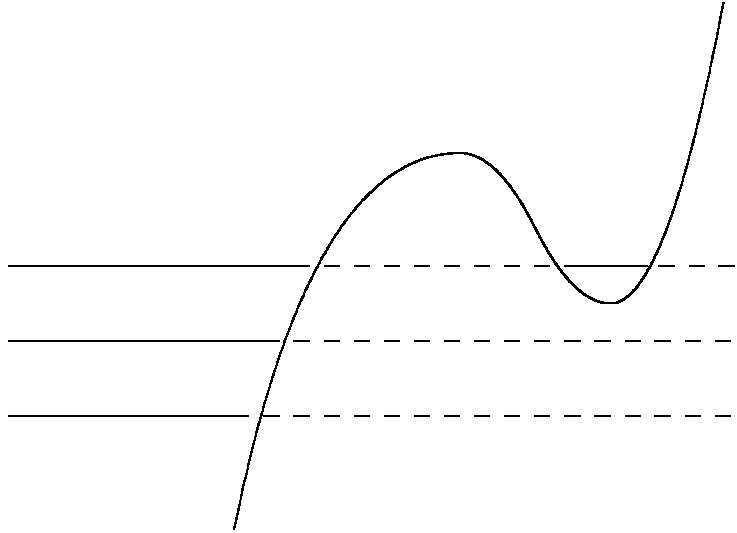
Fig.3 The energy levels in the two wells. The lowest energy in well
to the right coincides with the Fermi energy in the well to the
left if this energy is calculated according to the WKB prescription.

\vspace{12pt}

Until now we have determined
the Fermi energy by \rf{*I3}, which in scaled variables reads
\beq{*scfer}
\hbar N = \frac{1}{\pi} \int^{z_r(e_f)}_{z_l(e_f)} dz \; \sqrt{e_f-v_{fp}(z)}.
\eeq
As remarked above (see \rf{*I12}) this implies that $e_f=-2$, i.e. $e_f$
coincides with the local minimum of the right well. However, to be precise
one should in the context of WKB expansion replace $N$ by $N+\oh$ and
this leads to a shift of the Fermi energy such that it agrees with the
lowest energy in the right well.  We have now a situation identical to
the standard double well. By tunneling the degenerate energy levels split
in two and the lowest one will now have to be identified with the Fermi
energy. The shift is non-perturbative in $\hbar$ and is given by the
standard WKB formula:
\beq{*J5g}
e_f \to e_f  - \frac{\om_f}{\pi}\cdot e^{-\Gamma/\hbar}
\eeq
where the tunneling amplitude $\Gamma$ is given by
\beq{*J5f}
\Gamma = \int^{\tilde{z}_l}_{z_r} dz\;\sqrt{ v_{fp} (z) -e_f}
\approx  \int^1_{-2} dz\;\sqrt{v_{ds}(z) +2}
= \frac{12\sqrt{3}}{5}.
\eeq
In \rf{*J5f} $\tilde{z}_l$ is the position of the left turning
point for a classical particle with energy $e_f$ in the right well,
while $z_r$ as usual denotes the right turning point for a classical motion
with energy $e_f$ in the left well. Again we see that it is possible
in this semiclassical
calculation to replace $v_{fp}$ by $v_{ds}$, in which case the
integral can be performed since $v_{ds}(z)+2=(1-z)^2(2+z)$

The lower levels $e_n,~n > 0$
will receive contributions which are exponentially small in $1/\hbar$ compared
to the one received by $e_f$ since they are located below
the local minimum of the well to the right.
We note that the contribution \rf{*J5g}
is identical to the ``non-perturbative'' ambiguity which comes
from the Painleve equation of ordinary gravity (but in the stochastic
approach it is of course not an ambiguity). This observation was first
made in refs. \cite{spain,russians} and is in agreement
with the fact that the WKB expansion of the stochastically regularized model
agrees with the genus expansion of the original matrix model of pure
2d gravity.

\section{The strong coupling expansion}

The theory outlined above offers the possibility of a convergent {\it strong
coupling} expansion. It might be useful to recall a similar situation
in the case of the ordinary anharmonic oscillator in quantum mechanics.
Let the hamiltonian be given by
\beq{*S1}
H = -\frac{d^2}{dx^2} + x^2 + g x^4, ~~~~~~g > 0.
\eeq
The ordinary perturbation expansion around the harmonic oscillator
is only an asymptotic expansion. The reason is that $g\, x^4$ is {\it not}
a small perturbation for any value of $g > 0$ and clearly $g \to -g$
changes drastically the nature of $H$. For this reason the expansion
in powers of $g$ is  not a convergent power expansion, but only an asymptotic
expansion. It is, however, possible to analyze the {\it strong coupling}
region $g \to \infty$ by a simple scaling argument:
\beq{*S2}
H(x) = \l^{-1/2} \tH (y),~~~~~y=g^{1/6} x,~~~\l=\frac{1}{g^{2/3}}
\eeq
\beq{*S3}
\tH (y) = -\frac{d^2}{dy^2}+y^4 + \l y^2
\eeq
{}From this we deduce that $\tH(y)$ has a strong coupling expansion,
which is analytic in
$\l = 1/g^{2/3}$. In fact the potential $V(y)=  y^2$ satisfies the standard
requirement for being an analytic perturbation of
$$H_0 (y) = -\frac{d^2}{dy^2}+y^4$$
namely that $V$ is $H_0$-bounded, i.e. (1) the domain
$D(V) \supseteq D(H_0)$ and (2) $||V \psi ||_2 \leq a || H_0 \psi ||_2+
b|| \psi ||_2$ for some $a$ and $b$ and all $\psi \in D(H_0)$.
For an eigenvalue
$E_n$ of the original hamiltonian $H$ we can therefore write:
\beq{*S4}
E_n(g(\l)) = \l^{-1/2}\sum_{k=0}^\infty c_{nk}\l^{k}
\eeq
where the power series has a finite radius of convergence.

\vspace{12pt}

\noindent
The situation is very similar in the case of the hamiltonian $h_{fp} (z)$.
In sect. 4 we introduced a scaling which was designed to make contact
with the asymptotic genus expansion of ordinary 2d quantum gravity,
i.e. the limit $\hbar \to 0$. But  we can perform the same scaling argument
as just outlined for the anharmonic oscillator and derive an expansion
for {\it large} $\hbar^2$, i.e. in the strong coupling regime. If we
introduce scaled variables
\beq{*S5}
z=\hbar^{2/5} y,~~~~~\l=\frac{1}{\hbar^{4/5}}=\frac{g-g_c}{g_c}
\left(\frac{N}{2} \right)^{4/5}
\eeq
we can write
\beq{*S6}
h_{fp} (z) = \l^{-3/2}\tilde{h}_{fp}(y)
\eeq
\beq{*S7}
\tilde{h}_{fp}(y) = -\frac{d^2}{dy^2}+ \vt_{fp}(y,\l,N)
\eeq
\bea
\vt_{fp}(y)&=&y^3-3\l y +
\left(\frac{2}{N}\right)^{2/5}
\left[-\frac{3\l}{2} y^2 +\oq y^4\right]\nonumber \\
&=& \vt_{ds}(y,\l) + \left(\frac{2}{N}\right)^{2/5}
\left[-\frac{3\l}{2} y^2 +\oq y^4\right] \label{*S8}
\eea
and the double scaling limit is obtained as before for $N \to \infty$,
$g \to g_c$, but $\l$ fixed.

The non-trivial information is contained in the Hamilton function
$\tilde{h}_{fp} (y)$ which in the double scaling limit goes to
\beq{*S9}
\tilde{h}_{ds}(y) = -\frac{d^2}{dy^2}+\vt_{ds} = -\frac{d^2}{dy^2} +y^3-3\l y
\eeq
and we have formally the same situation as  for the anharmonic oscillator:
$-3\l y$ looks like a small perturbation with respect to $y^3$.
It is thus natural to expect a {\it strong coupling} expansion of
the energy eigenvalues of $h_{ds} (z)$ of the form:
\beq{*S10}
e_n (\hbar) = \l^{-3/2} \sum_{k=0}^\infty c_{nk} \l^k =
\hbar^{6/5} \sum_{k=0}^{\infty} c_{nk} \hbar^{-4k/5}
\eeq
where the series is convergent.  The domain of the perturbation $v(y) =
-3\l y$ does not include the domain of $h_0(y)= -d^2/dy^2 + y^3$  so
we have no rigorous proof  of this conjecture, but one should keep
in mind that the requirements (1) and (2) mentioned above are
 only  sufficient conditions, not  necessary conditions,
for analyticity.

\section{Discussion and conclusion}

We have shown that many of the results of the matrix models are
easily and transparently derived by means of stochastic quantization.
The asymptotic expansion, called the double scaling limit, is nothing
but the WKB expansion of the Fokker-Planck hamiltonian. In addition
stochastic quantization provides a non-perturbative definition of
2d quantum gravity. To be entirely correct one should first calculate
expectation values of observables using the full Fokker-Planck
hamiltonian and afterwards take the double scaling limit
$N \to \infty$, $g \to g_c$, with $\hbar \sim N^{-1} (g-g_c)^{-5/4}$ fixed.
If we reverse the procedure the double scaling limit of the Fokker-Planck
hamiltonian results in a hamiltonian $h_{ds}$ which is unbounded
from below and which has a one-parameter family of self-adjoint
extensions. One member is picked out by the requirement that the Fermi
energy should coincide with the corresponding energy of the full
Fokker-Planck potential. We mentioned  the possibility of a strong
coupling expansion, similar to the one of the anharmonic oscillator.

It is interesting to compare the situation described above  with the
similar situation present for 2d gravity coupled to a scalar field
(i.e. $c=1$). In this case we again have a matrix model description
which, in the double scaling limit,
results in a Schr\"{o}dinger eigenvalue equation:
\beq{*T1}
\left[ -\hbar^2\frac{d^2}{dy^2} - y^2 \right] \psi_n = \ep_n \psi_n .
\eeq
where $\hbar^{-1}= N (g_c -g)$ is fixed \cite{bkz,mg1,gz}.
Again the ground state is
fermionic and given by the Slater determinant of the $N$ eigenfunctions
counted from the top of the upside-down quadratic potential in \rf{*T1}.
One could be tempted to apply the methods outlined above to
this hamiltonian, which is unbounded from below, but condition
\rf{*infty} is {\it not} satisfied for this potential, and the problem
is classical complete: It takes an infinite time for a classical
particle to escape to infinity. In the terminology of Weil it is
a point-limit case of the Sturm-Liouville   theory of second order
differential equations. The point spectrum is empty and there is
a unique self-adjoint extension. If we cut off the  potential
at $\pm L$ we get a sequence of Hamiltonians which are bounded
from below and which convergences to the  self-adjoint extension of
the lhs of \rf{*T1} for $L \to \infty$.
The (generalized) eigenfunctions are parabolic cylinder
functions and they allow for the calculation of both local \cite{ajk} and
non-local \cite{moore} observables within the WKB approximation, but
although the self-adjoint extension of hamiltonian given by \rf{*T1}
is unique and defined beyond the WKB expansion, it is still not clear
whether it allows a strong coupling expansion for small $\hbar$ as
conjectured in \cite{mg1}.

\vspace{12pt}

\noindent
{\bf Acknowledgement.}
It is a pleasure to thank Tim Morris for discussions.

\addtolength{\baselineskip}{-0.20\baselineskip}

\end{document}